\documentclass[twocolumn]{aastex7}

\usepackage{amsmath}
\graphicspath{{./}{figures/}}
\usepackage{threeparttable}
\usepackage{booktabs}

\shorttitle{Solar Surface Flux Transport and Magnetic Power Spectra}
\shortauthors{Yukun Luo et al.}

\begin{document}

\title{Simulation of Solar Surface Flux Transport Constrained by Magnetic Power Spectra. \uppercase\expandafter{\romannumeral1}. Flux Transport Parameter}

\author{Yukun Luo}
\affiliation{School of Space and Earth Sciences, Beihang University, Beijing, People’s Republic of China}
\affiliation{Key Laboratory of Space Environment Monitoring and Information Processing of MIIT, Beijing, People’s Republic of China}
\email{luoyukun@buaa.edu.cn}

\author{Jie Jiang}
\affiliation{School of Space and Earth Sciences, Beihang University, Beijing, People’s Republic of China}
\affiliation{Key Laboratory of Space Environment Monitoring and Information Processing of MIIT, Beijing, People’s Republic of China}
\email[show]{jiejiang@buaa.edu.cn}

\author{Ruihui Wang}
\affiliation{School of Space and Earth Sciences, Beihang University, Beijing, People’s Republic of China}
\affiliation{Key Laboratory of Space Environment Monitoring and Information Processing of MIIT, Beijing, People’s Republic of China}
\email{wangruihui@buaa.edu.cn}


\begin{abstract}
The multi-scale structure of the solar surface magnetic field is essential for understanding both the Sun’s internal dynamo processes and its external magnetic activity. The surface flux transport (SFT) model has been successful in describing the large-scale evolution of the surface field, but its ability to capture observed multi-scale features, quantified by magnetic power spectra, remains uncertain. Here, we evaluate the SFT model by comparing observed and simulated power spectra across a broad range of spatial scales and by analyzing the effects of key transport parameters. We find that the simulations reproduce the observed spectra well at spherical harmonic degrees $l\lesssim60$, but diverge progressively at smaller spatial scales $l\gtrsim60$. This divergence likely arises from the diffusion approximation used to model the random walk of supergranulation. Power at $20 \lesssim l \lesssim 60$ is primarily determined by the magnetic flux sources, while at $l \lesssim 20$, the spectra are more sensitive to transport parameters. The meridional flow profile, including its equatorial gradient, peak latitude, and polar distribution, along with turbulent diffusivity, has distinct impacts on the low-degree modes ($l \leq 5$). In particular, a comparison of the $l$=3 and $l$=5 multipoles strengths suggests that the poleward flow above $\sim\pm60^\circ$ latitudes is very weak. This study presents the first quantitative validation of SFT models using magnetic power spectra and provides new constraints on surface flux transport physics.
\end{abstract}

\keywords{\uat{Solar magnetic fields}{1503} --- \uat{Solar active regions}{1974}  --- \uat{Solar cycle}{1487}
  --- \uat{Solar meridional circulation} {1874}}

\section{Introduction} \label{sec:intro} 
The solar surface magnetic field exhibits complex, multi-scale structures, spanning from global scales to active region (AR) scales and down to smaller magnetic network and internetwork features \citep{deWijn2009, BellotRubio2019}. These structures play a crucial role in both the internal dynamo processes and the dynamic phenomena of the upper solar atmosphere.  Its lowest-order multipoles dominate the structure of heliospheric magnetic field and heliospheric current sheet \citep{Wang2011,Wang2014}. The magnetic power spectrum provides an effective tool for quantitatively characterizing the distribution of magnetic energy across different spatial scales \citep{Abramenko2001, Luo2023, Luo2024, Kishore2025}. Meanwhile, to understand the evolution of the Sun’s surface magnetic field, the surface flux transport (SFT) model remains the prevailing framework.

The SFT model describes the transport of the radial magnetic field across the solar surface by horizontal flows, namely differential rotation, meridional circulation, and the random walk driven by supergranulation. For a historical overview of the model’s development, see the review by \citet{Sheeley2005}. Despite its relative simplicity, the SFT model has successfully reproduced key observational features \citep[e.g.,][]{Cameron2010, Yeates2020}. For a comprehensive review, see \citet{Yeates2023}. In particular, the SFT model has proven to be a powerful tool for studying the distribution and evolution of large-scale magnetic fields in the solar polar regions \citep[e.g.,][]{Wang1989,Jiang2009, Petrovay2019, Yang2024, Yeates2025, Pal2025}.

Although current SFT models have been successful in simulating the large-scale evolution of the solar surface magnetic field, several model parameters remain poorly constrained by observations. These parameters generally fall into two categories: transport parameters and source parameters \citep{Jiang2014, Wang2017, Yeates2023}. Among the transport parameters, differential rotation is relatively well established, except for its precise profile in the polar regions. In contrast, the meridional flow is much more uncertain, even at the solar surface, due to its relatively weak amplitude. \cite{Petrovay2020} propose that the latitudinal gradient of the surface poleward flow on the equator, $\Delta \upsilon$, plays a crucial role in the SFT model. As summarized by Table 1 of \cite{Jiang2023}, past studies show that $\Delta \upsilon$ ranges from 0.4 to 1.4 ms$^{-1}$deg$^{-1}$, depending largely on the measurement technique. Moreover, different studies report varying peak latitudes and flow speeds. Whether the poleward flow extends all the way to the solar poles also remains an open question, as direct and reliable observations in the polar regions are still lacking. 

The random walk induced by supergranulation corresponds to the small-scale flow field in the SFT models. It is typically approximated as a diffusion process, as originally proposed by \cite{Leighton1964}. \cite{Devore1984} validated the approximation under the condition that the magnetic field's spatial scale is larger than the turbulence correlation length. However, the diffusion coefficients derived from various measurement methods are subject to significant uncertainty. \cite{Schrijver2008} summarize observationally inferred coefficients, which range from 110 to 600 km$^2$ s$^{-1}$. \cite{Cameron2011} report a narrower range of 110-340 km$^2$ s$^{-1}$ based on magnetic energy decay rates. These diffusivity values are widely adopted in SFT models. Apparently, the turbulent diffusion approximation fails to reproduce the clumping of magnetic flux along supergranular network boundaries presented in observed magnetograms. To improve the appearance of simulated maps, several approaches have been developed to better represent the small-scale flow. For example, \cite{Upton2014a, Caplan2025} resolve supergranular convective cells and model them directly through the advection term, which requires higher spatial and temporal resolution. Computationally cheaper methods were also introduced \citep[e.g.,][]{Worden2000,Schrijver2001}. 

Regarding the flux source parameters in the SFT model, most previous studies simplify ARs as bipolar magnetic regions (BMRs). However, recent progress has highlighted the limitations of this approximation. \cite{Jiang2019} demonstrate that the evolution of complex ARs differs significantly from that of idealized BMRs. Several investigations confirm that the BMR approximation tends to overestimate the axial dipole strength through the whole solar cycle compared with a realistic configuration \citep{Yeates2020, Wang2021, Wang2024}. Furthermore, the significant influence of a few rogue ARs on solar cycle evolution has been increasingly recognized by the community \citep[e.g.,][]{Jiang2015,Nagy2017}. 


To address uncertainties in these transport parameters, several studies have attempted to optimize them using observational constraints from limited large-scale magnetic features, such as polar fields, and magnetic butterfly diagrams \citep{Lemerle2015, Whitbread2017, Petrovay2019, Dash2024}. However, there has been no effort to connect SFT models with multiscale features of the magnetic field, as shown by magnetic power spectra. Furthermore, comparisons between SFT outputs and observed magnetic power spectra offer a quantitative approach to evaluating the validity of the turbulent diffusion approximation across different spatial scales, although it is regarded as effective for the evolution of the magnetic field on spatial scales larger than supergranulation \citep{Leighton1964, Devore1984}. \cite{Luo2023, Luo2024} provide continuous magnetic power spectra over solar cycles 23 and 24, based on synoptic maps from Solar Dynamic Observatory (SDO)/Helioseismic and Magnetic Imager (HMI) and Solar and Heliospheric Observatory (SOHO)/Michelson Doppler Imager (MDI). These enable the investigation of the SFT model through magnetic power spectra.

In the series of studies, we constrain solar SFT simulations by comparing magnetic power spectra across different scales. In the first paper, we compare the simulated magnetic power spectra with observations by assimilating identified ARs from magnetograms into our SFT model, and investigate the influence of transport parameters on the power spectra. In the second paper, we will evaluate the impact of source parameters, especially the BMR approximation, on the power spectra.

This paper is organized as follows. We introduce the SFT model used in our study, including the data assimilation method and numerical treatment in Section \ref{sec:model}. In Section \ref{sec:results}, we validate our SFT model, evaluate the validated spatial scales for the turbulent diffusion, and investigate the influence of transport parameters on spectra. Section \ref{sec:conclusion} summarizes the results and gives the discussion.

\section{Surface Flux Transport Model}\label{sec:model}
The equation we use to describe the evolution of the large-scale magnetic field corresponds to the radial component of the magnetic induction equation, under the assumption of zero radial diffusion, as detailed reviewed by \cite{Yeates2023}. It is

\begin{equation}\label{eq1}
\frac{\partial B}{\partial t}+\nabla\cdot(U_sB) = \eta\nabla_s^2 B+S(\theta, \phi, t),	
\end{equation}	
where $B(\theta,\phi,t)$ is the surface radial magnetic field at co-latitude $\theta$, longitude $\phi$, and time $t$ in spherical coordinates, $U_s$ is the large-scale axisymmetric surface flow field including the differential rotation $\omega(\theta)$ and the meridional flow $v(\theta)$, $\nabla_s^2$ is the Laplacian operator on the surface of a sphere. The small-scale flow field due to the random walk of granulation and supergranulation is approximated as the turbulent diffusion \citep{Leighton1964} with the diffusivity value of $\eta$. The term $S(\theta, \phi, t)$ represents the emergence of the flux source due to radial flow on the horizontal magnetic field. The additional radial decay in the SFT equation proposed by \cite{Schirijver2002} and \cite{Baumann2006} are not considered here.  The details of the transport parameters and the method to deal with the source of magnetic flux are presented in the following two subsections, respectively.

\subsection{Transport parameters} \label{subsec:trans para}
For differential rotation $\omega(\theta)$ in this paper, we adopt the following formulation, which is defined with respect to the Carrington frame of reference:
\begin{equation}\label{eq6}
	\omega(\theta)=13.38-2.30 \cos ^{2} \theta-1.62 \cos ^{4} \theta -13.2
\end{equation}
in units of degrees per day \citep{Snodgrass1983}.

\begin{figure}[!h]
    \centering
    \includegraphics[width=1.0\linewidth]{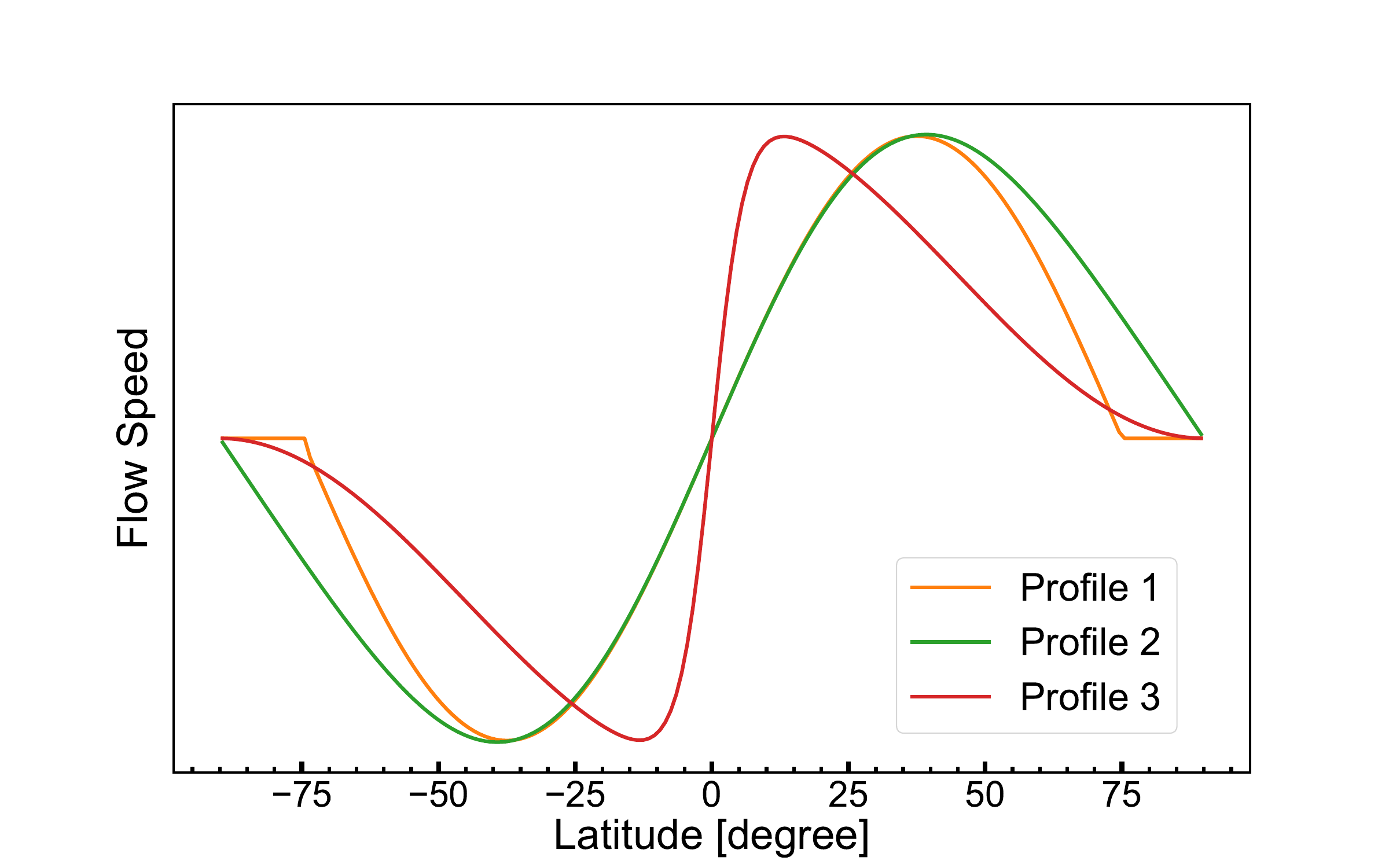}
    \caption{Latitudinal dependence of the three meridional flow profiles considered in this study. Their amplitudes vary between cases and are specified in the main text.}
    \label{fig1}
\end{figure}

The accurate meridional flow profile is controversial. In this paper, we adopt three different profiles to investigate their effects on simulated results. Profile 1 is proposed by \cite{vanBallegooijen1998}
\begin{equation}\label{eq2}
v(\lambda) = 
\begin{cases}
-v_{1} \sin \left( \pi \lambda / \lambda_{0} \right) & : \text{if } \left| \lambda \right| < \lambda_{0}, \\
0 & : \text{otherwise,}
\end{cases}
\end{equation}
where $v_1$ is the peak meridional flow speed, $\lambda$ is the latitude, and $\lambda_0$ is the cutoff latitude above which the flow vanishes. We adopt $\lambda_0=75^\circ$ so that the flow speed drops to zero above 75$^\circ$. This profile is adopted as the default in the subsequent analysis, as it has been widely used in recent studies. However, this does not imply that we consider it the most realistic profile. Profile 2 is a modified version of the one derived by \cite{Komm1993} based on magnetic feature tracking:
\begin{equation}\label{eq4}
	v(\lambda)=-v_{2} \left(31.4 \sin \lambda-11.2 \sin ^{3} \lambda\right) \cos \lambda .
\end{equation}
Comparing to Profile 1, the flow speed drops to zero at the poles and the peak flow speed determined by $v_{2}$. Both profiles peak at approximately $\pm$40$^\circ$ latitudes. In contrast, Profile 3 peaks at a low latitude ($\sim$13$^\circ$) and is much slower at higher latitudes than the other two profiles:
\begin{equation}\label{eq3}
    v(\lambda)=-v_{3} \tanh \left(\lambda / 6^{\circ}\right) \cos ^{2} \lambda .
\end{equation}
The latitudinal dependence follows the functional form given by \cite{Wang2017}, and $v_{3}$ determines the peak flow speed. The latitudinal dependence of the three meridional flow profiles are shown in Figure \ref{fig1}. Their amplitudes vary between cases and are specified below.

As the magnetic diffusivity $\eta$ is divergent in previous studies, we allow it to vary from 100 to 550 km$^2$ s$^{-1}$ in this study. The latitudinal flux transport process is influenced by both meridional flow and supergranulation diffusion. \cite{Petrovay2020} introduce a new parameter, ``dynamo effectivity range" $\lambda_R$, to quantify the combined effect of $\eta$ and the divergence of meridional flow at the equator $\Delta v$ on the axial dipole strength:
\begin{equation}\label{eq7}
    \lambda_{R}=\sqrt{\frac{\eta}{R_{\odot} \Delta v}}.
\end{equation}
In the following simulations, we use the first meridional flow profile with $v_1=13$ m s$^{-1}$ and $\eta=280$ km$^2$ s$^{-1}$, resulting in  $\lambda_R = 6.5^\circ$, as the default set of parameters.

In order to keep a constant value of $\lambda_R$ in the following comparison of meridional flow profiles, we set the the flow divergence at the equator $\Delta v$ to be the same for all three profiles, with $\Delta v$ = 0.545 m/(s$\cdot$deg), which is close to the value measured by magnetic feature tracking \citep{Komm1993, Hathaway2011}. 
Still, this value has been widely used in recent studies \citep{Jiang2023}, but we do not regard it as the most realistic. In the subsequent other analysis, we vary $\Delta v$ by adjusting the peak flow speed of $v_0$.  

\subsection{Flux source term $S(\theta, \phi, t)$} \label{subsec:source}
\begin{figure}[!h]
	\centering
	\includegraphics[width=1.0\linewidth]{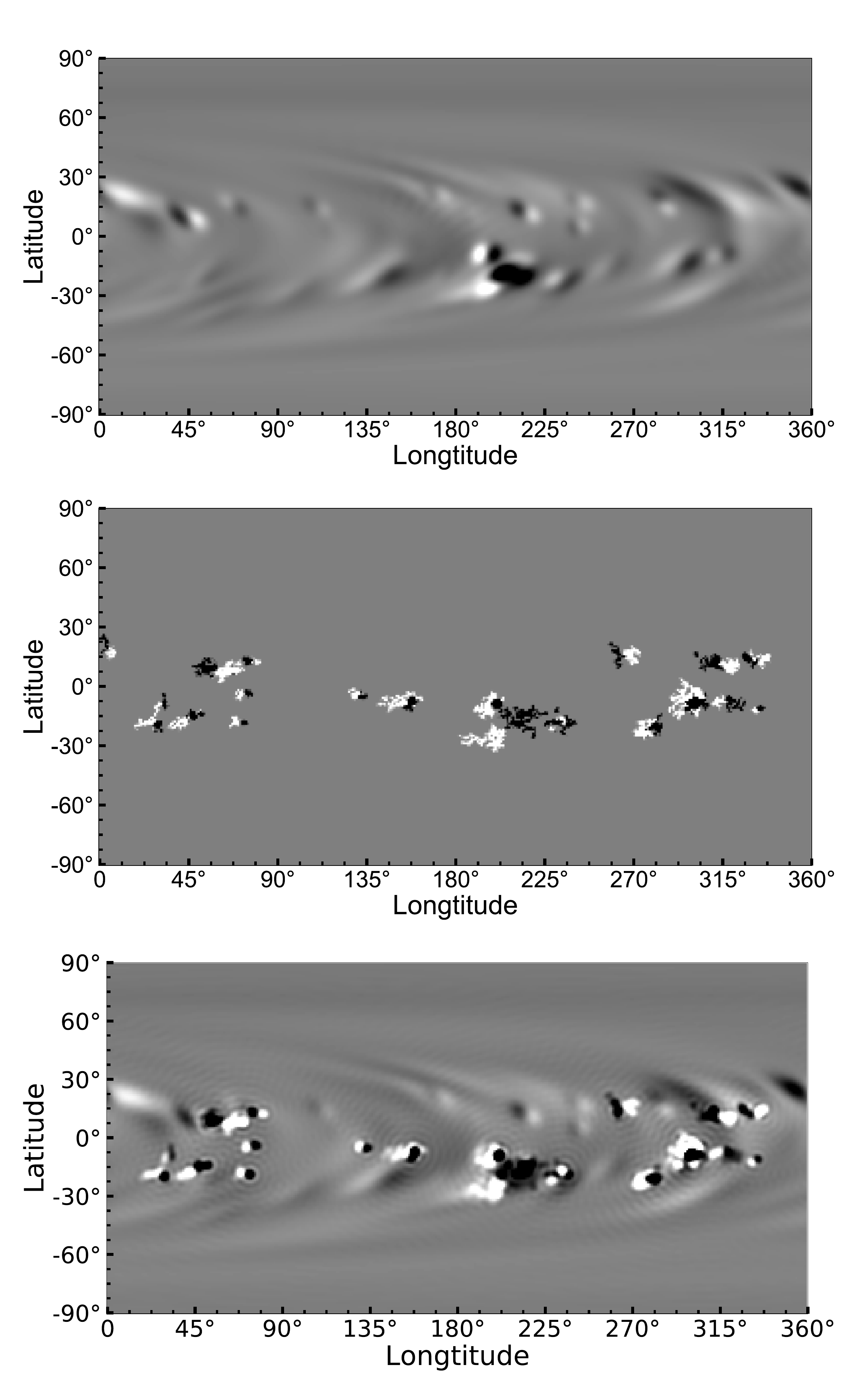}
	\caption{Illustration of magnetic flux source assimilation for the SFT simulation. The assimilation of the CR 1993 synoptic map is taken as an example. Upper panel: Magnetic field distribution from the SFT simulation on the last day of CR 1993, representing the final time step before flux assimilation. Middle panel: ARs identified from the CR 1993 synoptic magnetic map using the ARISES database. Lower panel: Updated magnetic field, in which the field from the upper panel is replaced by the detected ARs shown in the middle panel. The rings around ARs are the “ringing” artifact explained in Section \ref{subsec:numerical}}
	\label{fig2_fluxAssimilation}
\end{figure}

The flux source term $S(\theta, \phi, t)$ is constructed by assimilating individual ARs, detected from MDI synoptic maps, recorded in the Active Region database for Influence on Solar cycle Evolution (ARISE; \citealt{Wang2023, Wang2024}). The database and associated codes are publicly available on GitHub \footnote{\texttt{AR database:} \url{https://github.com/Wang-Ruihui/A-live-homogeneous-database-of-solar-active-regions}.} and version 3.0 is archived in Zenodo (\dataset[10.5281/zenodo.15076075]{https://doi.org/10.5281/zenodo.15076075}; \citealt{Database_WangRH_2025}). The ARISE database begins with Carrington Rotation (CR) 1909 and is continuously updated to the latest CR. For the objective of this paper, we only focus on cycle 23.   

Unlike previous flux assimilation techniques \citep[e.g.,][]{Yeates2015, Wang2025}, we do not assimilate ARs when they cross the central meridian. Instead, all ARs that emerge during the same CR are assimilated on the last day of that CR. This corresponds to an assimilation cadence of 1 CR, in contrast to the default 1-day time step used in the numerical algorithm described in Section \ref{subsec:numerical}. Figure \ref{fig2_fluxAssimilation} illustrates the process using the assimilation of ARs on the CR 1993 synoptic map as an example. The upper panel shows the magnetic field distribution from the SFT simulation at the final time step before flux assimilation, corresponding to the last day of CR 1993. The middle panel displays the ARs identified from the CR 1993 synoptic magnetic map using the ARISES database. After appropriate scaling presented below, all detected ARs are assimilated at once by replacing the field in the upper panel. The resulting magnetic field after assimilation is presented in the lower panel, which serves as the input for the next time step in the numerical simulation. This simultaneous assimilation of the full synoptic map facilitates direct comparison between the power spectra of the simulated and observed synoptic maps. 

It is essential to maintain overall net flux balance during each assimilation step. We continue using Figure \ref{fig2_fluxAssimilation} as an example to illustrate how this is ensured. The numerical algorithm presented in Section \ref{subsec:numerical} is free of the net flux inherently, so the magnetic field shown in the upper panel is already flux-balanced. To generate the updated field in the lower panel, we first identify the replacement region in the upper panel based on the position of the assimilated AR shown in the middle panel. We then calculate the two quantities: the net magnetic flux, $\phi_{net}$, within the replacement region of the upper panel, and the unsigned total flux, $\phi_{uns}$, of the detected ARs in the middle panel. To preserve both quantities after assimilation for each local replacement region, we apply separate scaling factors, $S_P$ and $S_N$, to the positive and negative polarity pixels of the assimilated ARs. The scaling factors are calculated by the following formula:
\begin{equation}\label{eq8}
    S_P=\frac{\phi_{net}+\phi_{uns}}{2\phi_P}, ~S_N=\frac{\phi_{net}-\phi_{uns}}{2\phi_N},
\end{equation}
where $\phi_P$ and $\phi_N$ are the flux for the positive and negative pixels of the assimilated AR, respectively.
Finally, the scaled ARs replace the corresponding region in the simulated map, ensuring that the overall net flux remains balanced in the lower panel of Figure \ref{fig2_fluxAssimilation}.

\subsection{Numerical Methods}\label{subsec:numerical}
To solve Equation (\ref{eq1}) on the solar spherical surface, we employ a spectral method with spherical harmonics as basis functions. The radial magnetic field $B(\theta,\phi,t)$ is expressed as
\begin{equation}\label{eq:SH_exp}
	B(\theta,\phi,t)=\sum_{l=0}^{\infty }\sum_{m=-l}^{l}  B_{l,m}(t) Y_{l,m}(\theta ,\phi ),
\end{equation}
where $Y_{l,m}(\theta, \phi)$ is the spherical harmonic function with degree $l$ and azimuthal order $m$, and $B_{l,m}$ is an expansion coefficient. The monopole solution $l=m=0$, which corresponds to the net magnetic flux over the solar surface and is incompatible with $\nabla\cdot B=0$, is excluded in the following simulation. 

This spectral approach provides several advantages. First, it inherently ensures that the magnetic field remains divergence-free on the surface and avoids singularities at the poles. These are particularly important because accurately capturing the polar field is a key objective of the SFT model. Even a small net flux introduced by numerical artifacts can lead to significant deviations in the polar field. 
Second, this method allows the diffusion term to be treated analytically as an eigenvalue problem of the spherical Laplace operator, which simplifies its numerical treatment. Third, it directly yields information about the evolution of magnetic multipoles, facilitating the analysis and comparison of power spectra.

The code was validated against the SFT code developed by \cite{Baumann2004}. Compared to their SFT code, one of the improvements is the use of the spherical harmonic iterative algorithm based on \cite{Holmes2002}, which enables higher computational efficiency and supports higher grid spatial resolution. The maximum spherical harmonic degree $l_{max}$ can be taken as at least 180 (corresponding to a spatial scale of 24 Mm). For the default case, we take it as $l_{max}=60$. 

Despite the advantages of the numerical methods presented above, the discontinuous flux source term illustrated in Figure \ref{fig2_fluxAssimilation} introduces the Gibbs phenomenon, manifesting as ringing artifacts in the magnetic field. These artifacts are visible in the lower panel of Figure \ref{fig2_fluxAssimilation} and Panels (c)-(d) of Figure \ref{fig4}. To mitigate these artifacts, we apply a Tukey window to $B_{l,m}(t)$ at a specific time $t$ when using Equation (\ref{eq:SH_exp}) to derive magnetograms.

The temporal discretization is performed using the fourth-order Runge-Kutta scheme. To ensure computational efficiency and numerical stability, the time step $\Delta t$ is set to about 1 day for the default case. For higher $l_{max}$ corresponding to higher grid resolutions, a smaller $\Delta t$ is required to maintain the stability of the simulation. 

The polar-corrected synoptic magnetograms observed by MDI/SOHO are used as the initial field for the numerical simulations. The original synoptic maps, with a resolution of 3600$\times$1440 pixels (uniform in longitude and in sine-latitude), are downsampled to 360 $\times$ 180 pixels (the default grid size), with uniform spacing in longitude and latitude. According to the sampling theorem, the maximum spherical harmonic degree $l_{max}$ should not exceed half the number of latitude grid points. Since this study focuses on cycle 23, we take the CR 1911 (1996 July) synoptic magnetogram as the initial condition of the subsequent simulations.

\section{Results} \label{sec:results}
\subsection{Validation of the model and the code}\label{subsec:validation}

\begin{figure*}[!ht]
    \centering
    \includegraphics[width=1.0\linewidth]{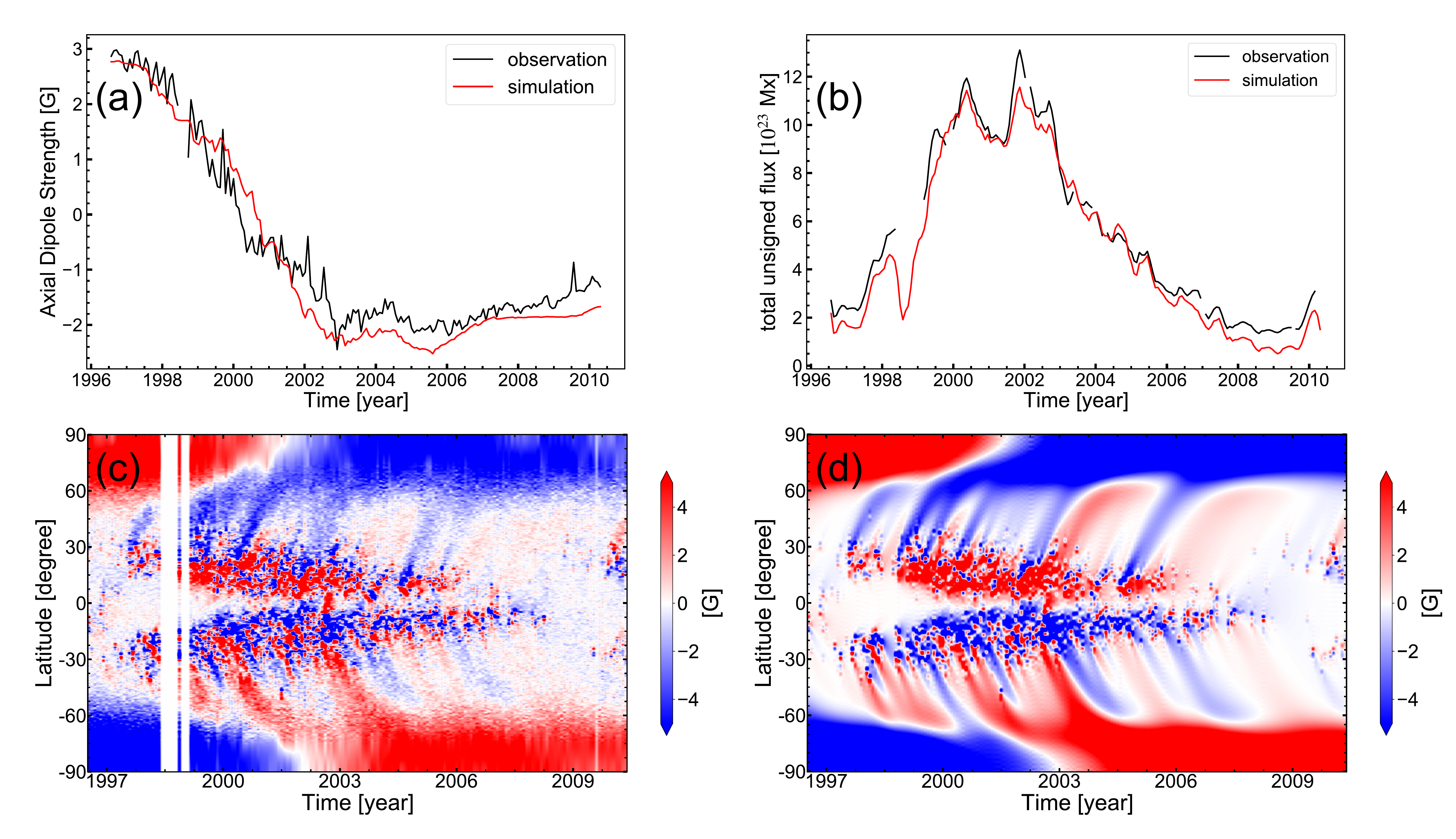}
    \caption{Comparison between observations and simulations. (a) Temporal evolution of axial dipole strength of the observation (black) and simulation (red). (b) Same as Panel (a) but for the total unsigned flux. (c) Time-latitude plot of the longitudinally averaged surface field, i.e., magnetic butterfly diagram, based on MDI synoptic maps. (d) Same as Panel (c) but derived based on simulated magnetograms.}
    \label{fig3}
\end{figure*}

To validate the reliability of our SFT model, we compare the simulated results using $l_{max}=60$ and $\Delta t=$ 1 day with observed ones. The simulated axial dipole strength, total unsigned flux, and magnetic butterfly diagram are shown in Figure \ref{fig3}. 

The temporal evolution of the axial dipole strength is shown in Figure \ref{fig3}(a). The results derived from the MDI synoptic maps (black curve) is calculated based on $\frac{3}{4\pi}\int_0^{2\pi}\int_0^{\pi}B(\theta,\phi,t)\cos\theta\sin\theta d\theta d\phi$, which is the spherical harmonic coefficient $B_{1,0}$. Accordingly, the simulated axial dipole strength (red curve) is directly given by the simulated result of $B_{1,0}$. The simulation shows a high degree of agreement with the observations throughout the entire solar cycle.

Figure \ref{fig3}(b) compares the total unsigned flux, defined as $R_\odot^2\int_0^{2\pi}\int_0^{\pi}|B(\theta,\phi,t)|\sin\theta d\theta d\phi$. When calculating the observed flux, we decompose the original observed map with $l_{max}=60$ to eliminate the effect of resolution differences. The abnormal decline in the simulation flux around 1999 is due to the lack of MDI data during this period, resulting in missing AR sources. Overall, the simulation agrees well with the observations during the cycle maximum, but underestimates the total unsigned flux during the cycle minima. As shown in Figure \ref{fig3}(a) and compared in Section \ref{subsec:power}, the simulated axial dipole strength matches the observations. Hence, the disagreement is unlikely to stem from the axial dipole field, approximately equal to the polar field, which is generally regarded as the dominant contributor to the total flux at cycle minimum. Instead, the discrepancy may be attributed to small-scale ephemeral regions \citep{Hofer2024}, which are not included in the data assimilation.

The temporal evolution of the longitudinal averaged surface field $B(\theta,\phi,t)$, i.e., the magnetic butterfly diagram, from observation and simulations is displayed in Panels (c) and (d) of Figure \ref{fig3}, respectively. They are in good agreement in terms of the polar reversal time and poleward plumes.

\begin{figure*}[!h]
    \centering
    \includegraphics[width=1.0\linewidth]{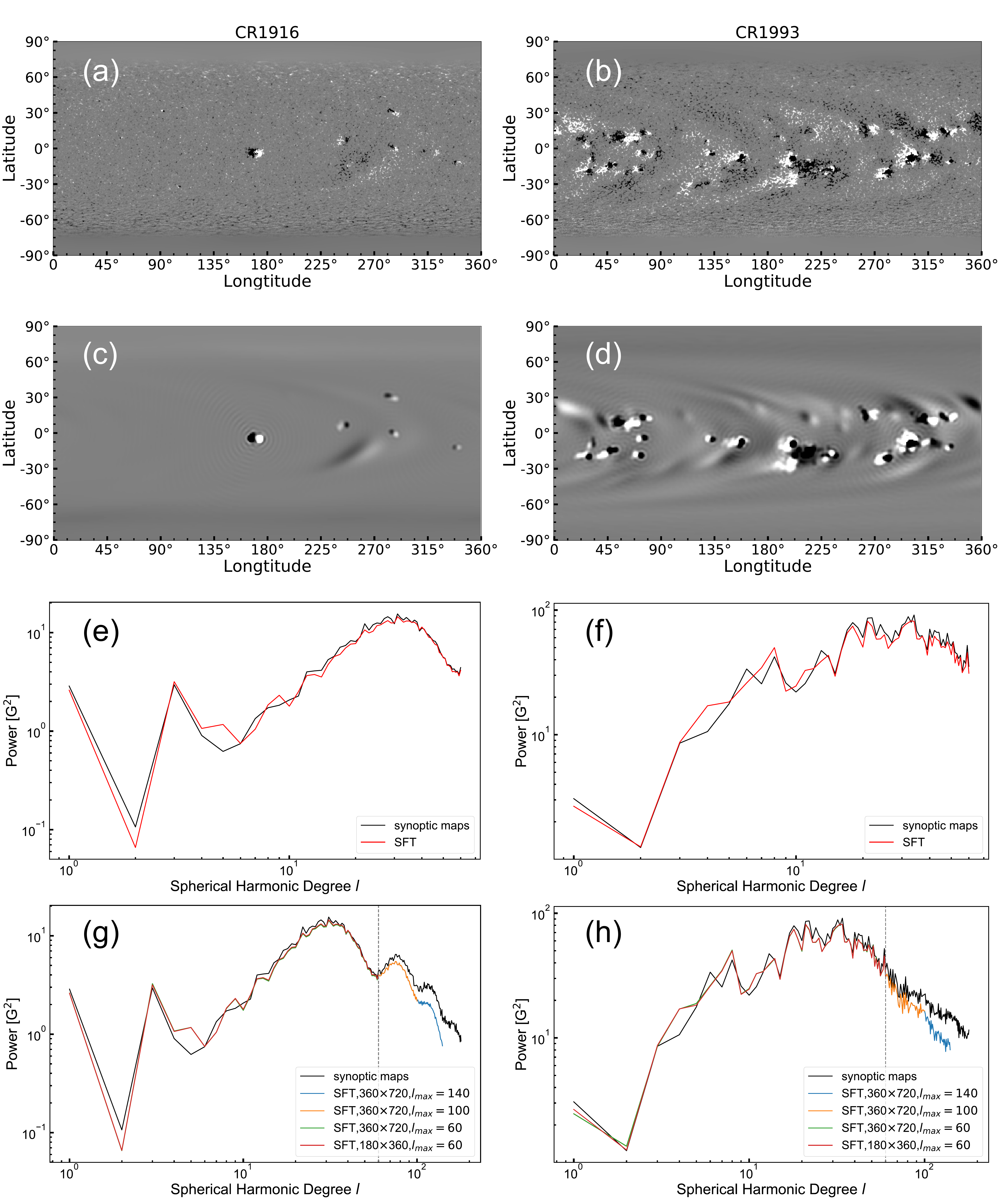}
    \caption{Comparison of synoptic maps and magnetic power spectra. CR 1916 (left) and CR 1993 (right) are taken as examples. (a)-(b) Synoptic maps, with magnetic fields saturated at 100 G. (c)-(d) Same as (a) and (b) but for the simulated magnetograms. The rings around ARs are the “ringing” artifact explained in Section \ref{subsec:numerical}. (e)-(f) Magnetic power spectra of synoptic maps (black, from (a)-(b)) and simulated magnetograms (red, from (c)-(d)), $l_{max}=60$. (g)-(h) Magnetic power spectra with $l_{max}=$60, 100, and 140 using two spatial resolutions of the initial magnetogram $180 \times 360$ and $360 \times 720$. The vertical gray lines mark the spatial scale at $l=$60.}
    \label{fig4}
\end{figure*}

We choose the observed and simulated synoptic maps of CR 1916 and CR 1993 as representative examples for further comparison. CRs 1916 and 1993 correspond to a solar minimum and a solar maximum, respectively. Their synoptic maps are displayed in Figures \ref{fig4} (a)-(d). The simulated maps exhibit large-scale magnetic structures that are generally consistent with those in the observations. A difference is the presence of ring-like artifacts around newly emerged ARs in the simulated maps (Panels (c) and (d)). These features arise from the spherical harmonics decomposition, which can introduce such artifacts when newly emerged ARs produce discontinuities in the magnetic field distribution, as discussed in Section \ref{subsec:numerical}.

All of the above comparisons validate the reliability of the SFT model and code, as they can produce reliable simulations consistent with observations. The good agreement also implies that the additional decay term of the SFT equation proposed by \cite{Schrijver2001} is not necessary when assimilating real AR configurations. This is supported by a similar result of \cite{Whitbread2017}, who get better agreement with the observed axial dipole strength by assimilating sources above the flux threshold rather than using idealized BMRs with a decay term.

\subsection{Comparison of the Magnetic Power Spectrum from Observations and SFT Simulations} \label{subsec:power}
In the previous subsection, we have validated the SFT model and the corresponding code by comparing several commonly used diagnostic parameters. In this subsection, we compare the magnetic power spectra derived from the observed synoptic map and the corresponding simulated magnetograms. The power spectra are computed following the method described in \citet{Luo2023, Luo2024}.

\subsubsection{Consistency Between Observed and Simulated Power Spectra for $l\leq$60 }
As presented in the previous subsection, the maximum spherical harmonic degree is set to $l_{max}=60$  in Equation (\ref{eq:SH_exp}) for the current simulation. Accordingly, we first compare the magnetic power spectra in the range $l=1\sim60$. Simulated and observed magnetograms of CRs 1916 and 1993 are used as representative examples. The corresponding power spectra are shown in Figures \ref{fig4} (e) and (f), respectively. Across the entire range, the simulated spectra (red curves) show good agreement with the observed ones (black curves). To quantify this agreement for $l\leq$60, we introduce metrics defined as follows. 

Since the power at different spherical harmonic degrees may span one or two orders of magnitude, we evaluate relative rather than absolute errors. We use the relative mean square error ($r_{MSE}$) to quantify the degree of consistency between observed and simulated power spectra at $l\leq60$ during single CR. The definition of $r_{MSE}$ is:
\begin{equation}\label{eq9}
    r_{MSE}=\frac{1}{60}\sum_{l=1}^{60}\frac{(P_{o,l}-P_{s,l})^2}{P_{s,l}^2},
\end{equation}
where $P_{o,l}$ and $P_{s,l}$ are the powers of observed and simulated magnetic power spectra at the same spherical harmonic degree $l$, respectively. A smaller $r_{MSE}$ indicates better agreement between the two spectra. We adopt a threshold of $r_{MSE}$=0.25 to determine whether the observed and simulated power spectra are consistent. Specifically, if $r_{MSE}\leq$0.25 for a given CR, we classify the pair as a consistent spectrum. To further evaluate the model's overall performance in reproducing observed spectra throughout cycle 23, we define the consistency percentage $f$, which is the ratio of the number of consistent spectra to the total number of 167 CRs within the cycle.

The $r_{MSE}$ values calculated from power spectra in Figures \ref{fig3} (e) and (f), along with the corresponding consistency percentage $f$, are shown in the second row of Table \ref{table}. For both CRs, the $r_{MSE}$ values are approximately 0.022, which is far below the threshold, further supporting the agreement between simulations and observations. The consistency percentage reaches up to 58.7 $\%$, indicating that the magnetic power spectra for most CRs are consistent with observations for $l\leq$60.

\begin{table*}[!ht]
    \centering
    \caption{Quantitative Comparison between Simulated and Observed Magnetic Power Spectra}
    \begin{threeparttable}
    \begin{tabular}{ccccc}
    \hline\hline
        ~ & ~ & $r_{MSE}$ for CR 1916 & $r_{MSE}$ for CR 1993 & Consistency percentage $f$ \\ \hline
        Resolution of initial map & 180*360 & 0.022 & 0.021 & 58.7\% \\ 
        ~ & 360*720 & 0.024 & 0.022 & 58.1\% \\ \hline
        $l_{max}$ & 60 & 0.024 & 0.022 & 58.1\% \\ 
        ~ & 100 & 0.024 & 0.022 & 58.1\% \\ 
        ~ & 140 & 0.024 & 0.022 & 58.1\% \\ \hline
        Transport parameter & $\eta$=250, $v_0$=11.8 & 0.020 & 0.024 & 51.5\% \\ 
        ~ & $\eta$=350, $v_0$=16.3 & 0.028 & 0.018 & 67.1\% \\ 
        ~ & $\eta$=500, $v_0$=23.2 & 0.041 & 0.022 & 68.3\% \\ \hline
        Meridional flow profile & Profile 1 & 0.022 & 0.021 & 58.7\% \\ 
        ~ & Profile 2 & 0.080 & 0.026 & 37.1\% \\ 
        ~ & Profile 3 & 0.008 & 0.032 & 31.7\% \\ \hline
    \end{tabular}
    \end{threeparttable}
    \label{table}
\end{table*}


\subsubsection{Discrepancy at small scales $l>$60 and Independence from Truncation Degree $l_{max}$}
To examine the magnetic power spectra at smaller spatial scales, we increase the maximum spherical harmonic degree $l_{max}$. Accordingly, this requires increasing the spatial resolution of the initial synoptic maps. Thus we consider three cases with $l_{max}=60$, 100, and 140, all based on the initial map with the resolution of $360\times720$. Figures \ref{fig4} (g)-(h) compare the spectra over the range $l=1\sim140$ for the CRs 1916 and 1993 maps, respectively. Although the simulated spectra for $l_{max}=60$, $l_{max}=100$, and $l_{max}=140$ are plotted in different colors, they are nearly indistinguishable from each other for $l\leq$60. This is further confirmed by the identical $r_{MSE}$ and consistency percentage $f$ values listed in Table \ref{table}. These results not only demonstrate the convergence of the code but also reinforce the conclusion that the SFT simulations reliably reproduce the magnetic power spectra at large scales ($l\leq60$).

Although significant progress has been made in understanding the evolution of the solar surface magnetic field in recent decades, to our knowledge, this comparison represents the first successful reproduction of its evolution across a wide range of spatial scales. Additionally, the spherical harmonic treatment in numerical calculation of the SFT model offers high computational efficiency, enabling efficient large-scale simulations.

On the other hand, Figures \ref{fig4} (g)-(h) reveal a key discrepancy: at small scales ($l > 60$), the simulated spectra progressively deviate from observations, with the simulations exhibiting systematically lower power at higher harmonic degrees. According to \cite{Luo2024}, the observed magnetic power spectra display peaks or knees at scales of 26 Mm ($l=169$) to 41 Mm ($l=106$), corresponding to flux concentrations at supergranular cell boundaries. These features arise from the interaction between the magnetic field and supergranular flows, which are approximated as turbulent diffusion in our SFT model. The typical decay time $T_l$ of magnetic structures varies with the corresponding spherical harmonica degree $l$ with diffusion approximation \citep{Leighton1964}. The relation is
\begin{equation}\label{eq10}
	T_l = \frac{R_{\odot}^2}{\eta l(l+1)}.
\end{equation}
For example, $T_{60}=5.5$ days means that the magnetic structures with $l>60$ decay significantly within 5.5 days purely due to diffusion, rather than persisting through concentration at magnetic network boundaries. Furthermore, the structures smaller than ARs scale range, from $l=12$ (365 Mm) to $l=56$ (78 Mm), are not assimilated into the model. These factors collectively explain the weaker simulated spectra at scales $l>60$. The spectral deviations at the small scales thus reflect inherent limitations in the diffusion approximation’s ability to capture small-scale supergranular dynamics. But for the simulations of large-scale fields at scales $l\leq60$, our method remains both effective and computationally efficient.

\subsection{Effects of Transport Parameter}\label{subsec:effects}
We have presented the first successful comparison of magnetic power spectra over a wide range of scales between simulated and observed synoptic maps in the previous subsection. The evolution of the surface magnetic field is influenced by transport parameters such as the meridional flow and turbulent diffusivity $\eta$. Previous studies \citep[e.g.,][]{Lemerle2015, Virtanen2017, Whitbread2017, Yeates2023} have tested the influence of these parameters, aiming to constrain their values to better match observations. They found that the maximum meridional flow speed and diffusivity cannot be independently constrained. \citet{Petrovay2020} proposed a combined parameter, the dynamo effectivity range $\lambda_R$ (Equation~\ref{eq7}), which captures their joint influence. In this subsection, we evaluate the effects of transport parameters on magnetic structures at different spatial scales, considering cases with the same $\lambda_R$, varying $\lambda_R$, and different meridional flow profiles.

\begin{figure*}
	\centering
	\includegraphics[width=1.0\linewidth]{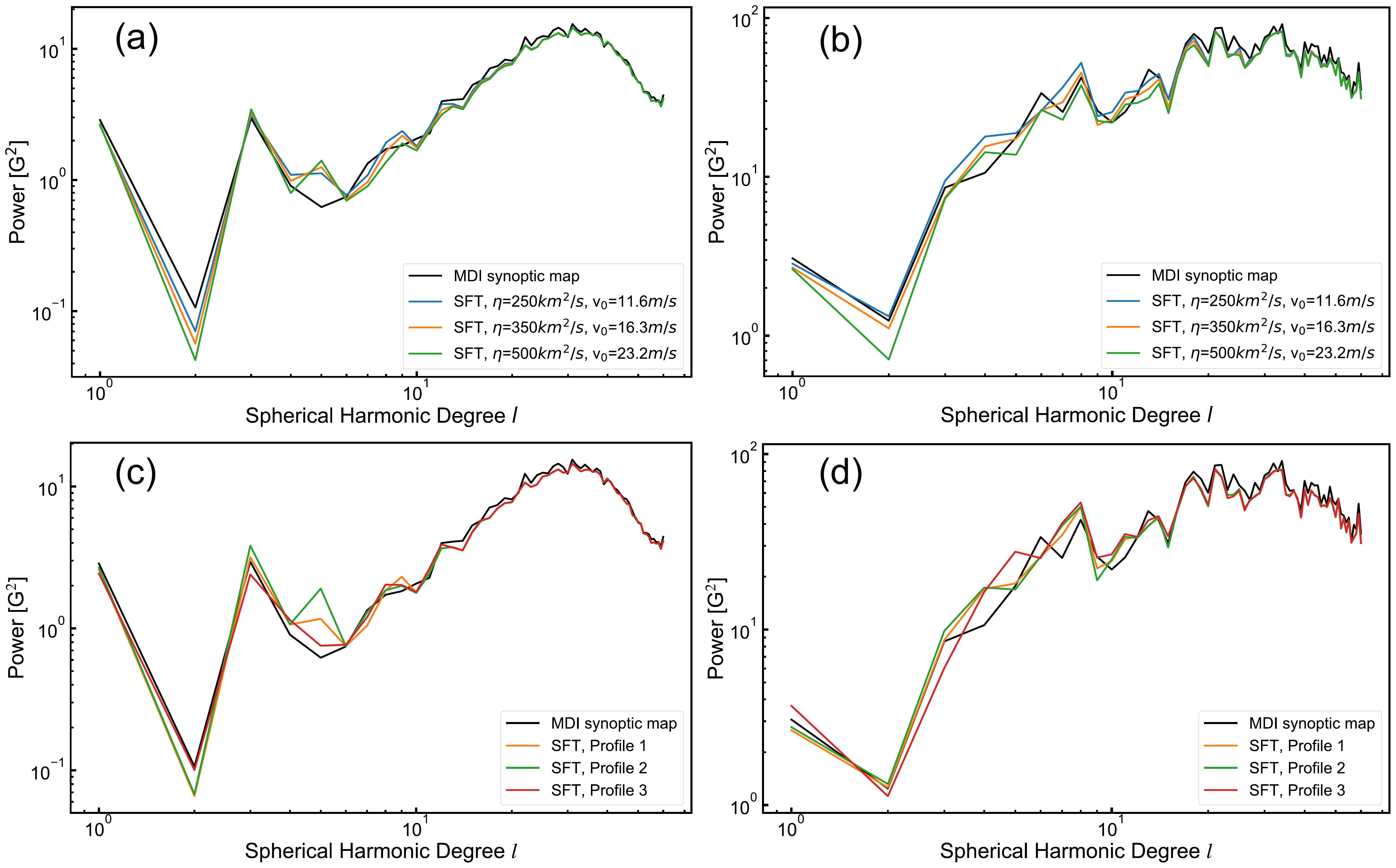}
	\caption{Examples of magnetic power spectra of CR 1916 (left) and CR 1993 (right), respectively. (a)-(b) The SFT models run with the same $\lambda_R$ but different diffusion coefficients $\eta$ and peak meridional flow speeds $v_0$. (c)-(d) The simulated power spectra, using the three different meridional flow profiles shown in Figure \ref{fig1}.}
	\label{fig5}
\end{figure*}

\subsubsection{Effects of Transport Parameters on Magnetic Spectra with Fixed $\lambda_R$}
\label{subsubsec:same}
We take the default meridional profile, i.e., Equation (\ref{eq2}), to carry out the study. The dynamo effectivity is fixed at $\lambda_R = 6.5^\circ$, and we vary $\eta$ among 250, 350, and 500 km$^2\cdot$s$^{-1}$. The corresponding $v_1$ values range from 11.6 to 23.2 m$\cdot$s$^{-1}$, resulting in $\Delta v =$ [0.49, 0.97] m/(s$\cdot$deg), which are within the value range measured by magnetic feature tracking and helioseismology \citep{Komm1993, Zhao2014, Mahajan2021,Jiang2023}. To illustrate the impact on magnetic power spectra, we still take CRs 1916 and 1993 as representative examples and compare the magnetic power spectra between simulated and observed maps. The results are shown in Figures \ref{fig5} (a) and (b).  

Although the overall spectra from simulations with the three parameter sets closely match the observations, noticeable differences appear at the low-$l$ (large-scale) end. This arises because different spectral ranges are shaped by different physical processes. For $20 \lesssim l \lesssim 60$, the power is primarily determined by the assimilated AR sources. Once ARs are accurately assimilated, the resulting spectra become consistent across different sets of transport parameters, explaining the similarity for $l \gtrsim 20$. In contrast, large-scale features ($l \lesssim 20$) are mainly shaped by the flux transport processes, which are governed by the transport parameters when the assimilated flux source is fixed. Furthermore, small-scale structures decay more rapidly due to turbulent diffusion. As a result, during the cycle minimum, spectral differences are mainly concentrated in the low-degree modes ($l \leq 5$), as presented in Figures \ref{fig5} (a). The figures also reveal distinct behaviors between even and odd $l$ modes, the origin of which is investigated below.
 
For a given $l$, the spectrum is calculated by combining contributions from all $m$ \citep[for details, see][]{Luo2023}. Among these, the axisymmetric modes ($m = 0$) are unaffected by differential rotation and therefore exhibit longer lifetimes. As a result, the $m = 0$ components dominate the power at each $l$ during the cycle minimum, and we focus exclusively on them in this analysis. The odd and even $l$ modes correspond to the anti-symmetric and symmetric modes, respectively. The poleward flow tends to balance the equatorward diffusion to establish the anti-symmetric time-asymptotic solution, corresponding to the odd $l$ modes. Hence the power in odd-$l$ modes (i.e., $l = 1$, 3, and 5) remains largely unchanged when varying $v_0$ and $\eta$ under a fixed $\lambda_R$. In contrast, increasing either $v_0$ or $\eta$ enhances the decay of symmetric modes, reducing the power in even-$l$ components. Hence, the power in the $l = 2$ and $l = 4$ modes decreases with increasing $v_0$ and $\eta$.

Figure \ref{fig5} (b) shows the spectral comparison for CR 1993, which corresponds to the cycle maximum phase. The distinction between even and odd modes persists but is less pronounced. During this phase, non-axisymmetric components ($m \neq 0$) from newly emerging ARs contribute significantly to the spectral power, resulting in behavior that differs from that seen during cycle minimum. 

The $r_{MSE}$ values for CRs 1916 and 1993 and consistency percentages $f$ for the whole cycle 23 are shown in Table \ref{table}. The $r_{MSE}$ values change slightly with the variation of parameters, but are still within the threshold range. Although during the cycle minimum, the set of parameters $v_0$=23.3 m$\cdot$s$^{-1}$ ($\Delta_v=0.972$m/(s$\cdot$deg)) and $\eta=500$ km$^2$$\cdot$s$^{-1}$ produce weaker symmetric modes than observations, the $f$ value is larger than the other two sets of parameters, indicating good consistency with the observed spectrum overall.

\subsubsection{Effects of Transport Parameter With Varied $\lambda_R$}\label{subsubsec:diff}

\begin{figure}
    \centering
    \includegraphics[width=1.0\linewidth]{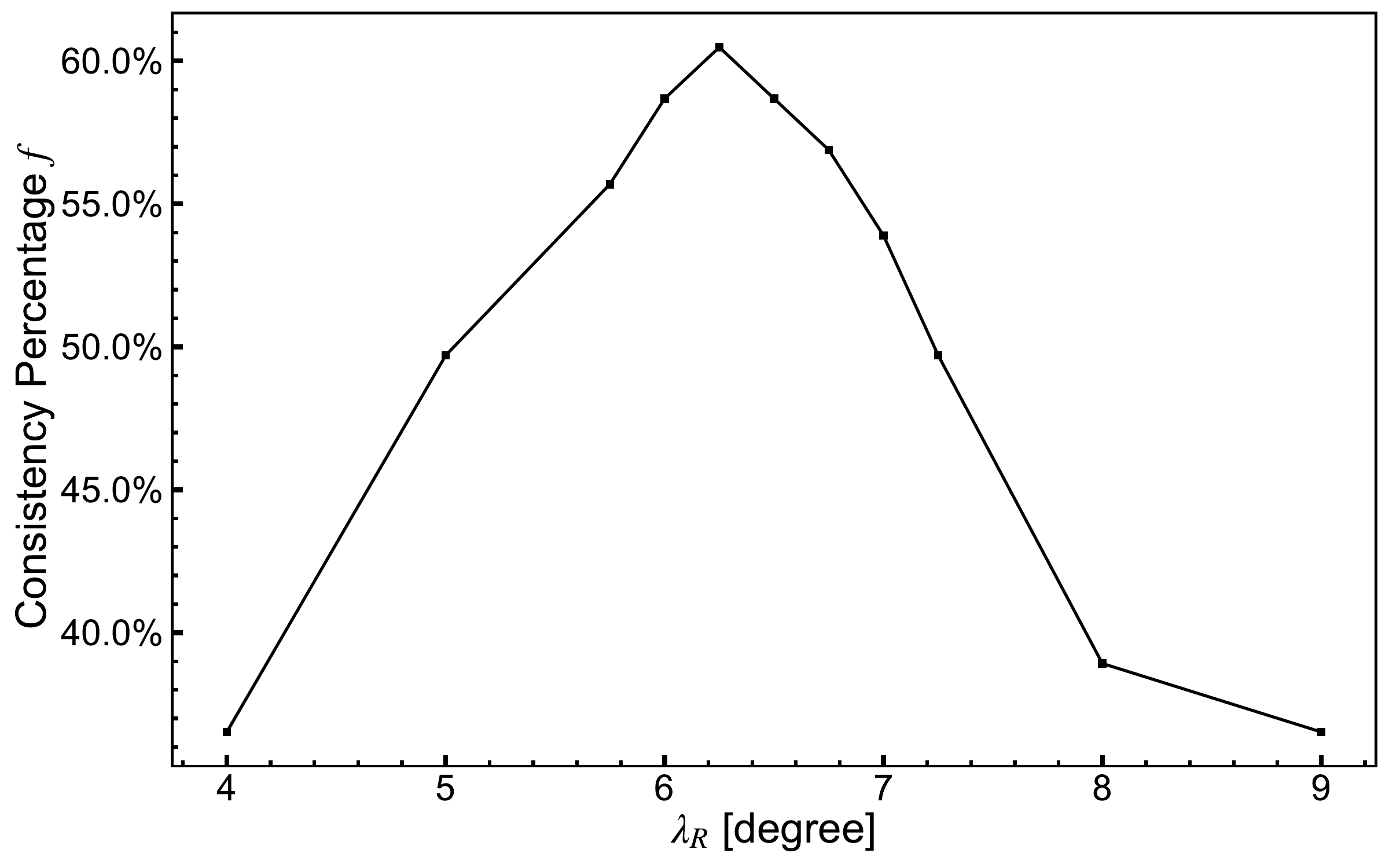}
    \caption{Variation of the consistency percentage $f$ with dynamo effectivity range $\lambda_R$.}
    \label{fig6}
\end{figure}

In this subsection, we further consider the effect of varied $\lambda_R$. We vary $\lambda_R$ from $4^\circ$ to $9^\circ$ by adjusting the diffusion coefficient from 106 km$^2$ s$^{-1}$ to 537 km$^2$ s$^{-1}$, while keeping $v_1=13$ m s$^{-1}$ in Equation (\ref{eq2}). 

Figure \ref{fig6} shows the variation of the consistency percentage $f$ with the dynamo effectivity range $\lambda_R$. As $\lambda_R$ increases, the $f$ values first rise and then decline, reaching a peak of 60.5\% at $\lambda_R = 6.25^\circ$. On either side of this peak, $f$ drops significantly, with a minimum of 36.5\% at both $\lambda_R = 4^\circ$ and $9^\circ$. Within the range $\lambda_R = 5.75^\circ$ to $6.75^\circ$, the $f$ values remain above 90\% of the peak value. Therefore, we suggest that transport parameters corresponding to this $\lambda_R$ range produce simulated magnetic power spectra that are in good agreement with observations when using Profile 1 of the flow.

Quantitatively, within our suggested range, the turbulent diffusion coefficient $\eta$ can vary from approximately 220 to 300~km$^2$ s$^{-1}$ when the peak meridional flow speed is fixed at $v_1 = 13$ m s$^{-1}$. Conversely, for a fixed $\eta = 500$ km$^2$ s$^{-1}$ suggested in Section \ref{subsubsec:same}, the corresponding $v_0$ values range from 21.6 to 29.6 m s$^{-1}$, with $\Delta v=$ [0.905, 1.24] m/(s$\cdot$deg). These parameter combinations are commonly adopted in previous studies, as summarized in Table 1 of \citet{Jiang2023}, and are also consistent with the parameter ranges suggested by \citet{Lemerle2015}. It is important to note that our constraint on $\lambda_R$ is based on Profile 1. Alternative flow profiles may lead to different optimal values of $\lambda_R$. Therefore, in the next subsection, we examine the effects of varying meridional flow profiles.

\subsubsection{Effects of Various Meridional Flow Profiles}\label{subsubsec:profile}

In this subsection, we investigate the effects of different meridional flow profiles on the magnetic power spectra using the three commonly adopted latitudinal dependences of the meridional flow shown in Figure~\ref{fig1}. For each profile, we fix the dynamo effectivity range at $\lambda_R = 6.5^\circ$ and ensure that the equatorial gradient of the flow, $\Delta v$, is the same across all cases, rather than matching the peak flow speed. As a result, the profiles reach their maximum speeds at different latitudes and exhibit different peak values. The default value of diffusivity $\eta=280$ km$^2$ s$^{-1}$ is used.

Figures \ref{fig5} (c) and (d) are the comparison between simulated power spectra using the three meridional flow profiles and observed ones at CRs 1916 and 1993. Considering the quantitative comparison shown in Table \ref{table}, Profile 1 performs better than the other two profiles. For $20 \lesssim l \lesssim 60$, the power is primarily determined by the assimilated AR sources. The effects of the three meridional flow profiles are presented in small $l$ modes. 

A prominent effect of the three meridional flow profiles on the magnetic power spectra at cycle minimum (Figure \ref{fig5}(c)) is observed in the $l=3$ and $l=5$ modes. Profiles 1 and 2 differ only in the polar regions, which likely accounts for their divergence at these specific modes. Profile 2, characterized by strong poleward flow near the poles, produces an $l=5$ power nearly three times stronger than observed. In contrast, Profile 1, which has zero flow above $\pm$75$^\circ$ latitude, generates weaker power at both $l=3$ and $l=5$. Profile 3 exhibits near-zero flow even above $\pm$60$^\circ$ latitude and produces the lowest power at these modes, most closely matching the observed spectrum. This comparison suggests the influence of polar flow on the power at $l=3$ and $l=5$. The underlying mechanism is analyzed below.

Here we still only analysis the axisymmetric modes which dominates the power at $l=3$ and $l=5$ at the cycle minimum. The spherical harmonics $Y_3^0$ and $Y_5^0$ have the nodes near $\pm$50$^\circ$ and $\pm$65$^\circ$ latitudes, respectively. This means that the spectral powers at $l=5$ and $l=3$ during cycle minimum dominate how magnetic flux is distributed near the poles. Larger powers at these modes correspond to stronger concentration near the poles, and vice versa. As demonstrated by \cite{Devore1984}, strong polar cap flows, especially Profile 2, tend to concentrate magnetic flux above $\pm$75$^\circ$ latitudes, whereas observations indicate that the polar field is concentrated above $\pm$60$^\circ$ as illustrated by Figure \ref{fig3}(c).  Thus, the spectral powers at $l=3$ and $l=5$ serve as effective diagnostics for constraining the latitudinal distribution of the meridional flow near the poles. The better consistency of the spectra at $l=3$ and $l=5$ for Profile 3 and the overestimate of the two modes for Profiles 1 and 2 during the cycle minimum support that the poleward flow should weaken significantly above $\pm$60$^\circ$ latitudes. A similar conclusion was reached by \cite{Lemerle2015}, who optimized meridional flow parameters to match the observed magnetic butterfly diagram of cycle 21. Figure 4 of \cite{Wang2017} also illustrated a similar scenario regarding the causal relationship between the polar distribution of the meridional flow and the magnetic field. 

Another notable effect of the flow profiles on the spectra at cycle minimum (Figures \ref{fig5} (c)) appears in the $l=2$ modes. Profile 3, which has the weakest peak flow speed $v_0$, produces the strongest power at $l=2$, in agreement with observations. In contrast, Profiles 1 and 2, both characterized by similarly stronger peak flow speed $v_0$, produce comparably weaker power at $l=2$. The weak flow in Profile 3 leads to a slow decay rate of the quadrupole modes, resulting in higher spectral power. However, for a CR map during the cycle 23/24 minimum, e.g., CR 2064, the weak decay cause the quadrupole power to exceed the observed value significantly. These results highlight the sensitivity of the $l=2$ mode to the peak meridional flow speed, which may influence the north-south asymmetry of solar activity \citep{Schussler2018}. The dominance of the $l=2$ mode during polar field reversal may give rise to a conical heliospheric current sheet \citep{Wang2014, Wang2014b, Khabarova2017}. 

The above analysis is based on the time-asymptotic solution of the SFT equation \citep{Devore1984, DeVore1987}, i.e., Equation (\ref{eq1}), which corresponds to a cycle minimum without new flux emergence. Among the three cases, Profile 3 reproduces the spectral power at low-degree modes ($l \leq 5$) during the cycle 22/23 minimum reasonably well. However, its performance during other solar cycle phases is worse, as presented in Figure \ref{fig5} (d). Each emerging AR contributes power, including both axisymmetric and non-axisymmetric modes at specific $l$, resulting in power levels more than an order of magnitude higher than those at solar minimum. The low peak latitude of Profile 3 results in a weaker amplitude when $\Delta v$ is kept the same as in the other two profiles. The $f$ value quantifying spectral consistency is 0.032, in contrast to 0.021 and 0.026 for Profiles 1 and 2, respectively. Table \ref{table} further indicates that the consistency fraction $f$ for Profile 3 is only 31.7$\%$ for the whole cycle 23, suggesting that most of the simulated magnetic power spectra are inconsistent with observations. If its amplitude is increased to 13 m s$^{-1}$ as in the other profiles, the corresponding latitudinal gradient $\Delta v = 2.17$ m s$^{-1}$ deg$^{-1}$ becomes at least twice the typical observational estimates \citep{Jiang2023}, resulting in insufficient cross-equatorial flux transport. Our tests confirm that under such conditions, the model fails to reproduce polar field reversal. 
In contrast, Profile 1, which has a stronger $v_0$ and a higher-latitude peak, produces a consistency fraction $f$ of 58.7\%. Their specific property of the meridional flow profile, especially the peak latitude of flow, remains to be investigated. 

\section{Conclusion and Discussion} \label{sec:conclusion}

In this study, we present the first attempt to link the SFT model, which is widely used to describe the distribution and evolution of the Sun’s surface magnetic field, with magnetic power spectra. By comparing spectra derived from simulated and observed magnetograms, we not only validate the SFT model beyond traditional diagnostics such as axial dipole strength, total unsigned flux, and the magnetic butterfly diagram, but also gain new insight into the distinct effects of transport parameters on large-scale power ($l \leq 5$). The strength and evolution of these lowest-order multipoles are particularly important for determining the heliospheric magnetic field strength and the position of the heliospheric current sheet.

Our comparisons show that the simulated power spectra generally agree well with observations for spherical harmonic degrees $l \lesssim 60$. However, at smaller spatial scales ($l \gtrsim 60$), the simulated spectra progressively deviate from the observed spectra, reflecting the inherent limitations of the diffusion approximation. Nonetheless, the good agreement at large scales demonstrates that the diffusion approximation remains effective for capturing large-scale magnetic features.

Although the turbulent diffusion approximation of supergranulation flow fails to reproduce the spatial scales smaller than 73 Mm ($l\gtrsim$ 60), it has a significant advantage in terms of computational efficiency for modeling the large-scale magnetic field. This efficiency is particularly notable in our implementation, where the diffusion term is treated as an eigenvalue problem. In all cases presented in this study, computing the diffusion term requires less than one-tenth of the total runtime. In contrast, finite-difference or finite-volume methods, such as the model of \citet{Caplan2025} designed to capture small-scale magnetic features, are computationally much more expensive. The effectiveness of such models in reproducing magnetic power spectra across a broad range of spatial scales remains to be systematically evaluated. 
Given the accuracy and high efficiency of our model in reproducing large-scale magnetic structures, the solar surface magnetic field evolution derived from our approach could provide valuable insight in understanding the polar field \citep{Wang2020,Yang2024}, solar open flux \citep{Wang2000, Fisk2001, Linker2017}, coronal structure \citep{Downs2025}, and the heliospheric magnetic field \citep{Owens2013}, which are planned for exploration in forthcoming studies.   

The spectral power at $l \lesssim 20$ is dominated by the transport parameters. We evaluate the impact of turbulent diffusivity $\eta$ and various meridional flow characteristics, including amplitude, equatorial gradient $\Delta v$, and polar distribution, on different modes of magnetic structures. When maintaining the same dynamo effectivity range, $\lambda_R$, and meridional flow profile, simulations with relatively larger values of $\eta$ and $\Delta v$ show slightly better agreement with observations. Through systematic tests of different $\lambda_R$ values, we identify an optimal range of $\lambda_R = 5.75^\circ$–$6.75^\circ$. Comparisons among three meridional flow profiles further highlight the critical influence of the peak speed and polar distribution of the meridional flow on the low-degree spectral features.

Our proposed optimal range is close to that of \cite{Yeates2025}, who optimize transport parameters using reconstructed historical data and suggest a best-fit value of $\lambda_R=5.94^\circ$, with an acceptable range of $\lambda_R = 6.58 \pm 0.99^\circ$. The optimization result of \cite{Wang2024} using an algebraic method is $\lambda_R = 5^\circ$ for MDI maps, which is slightly below our proposed range. \cite{Talafha2022} demonstrate that $\lambda_R<10^\circ$ means latitude quenching, i.e., the modulation of polar field generation arising from the cycle-dependent mean latitude of active region emergence \citep{Jiang2020, Petrovay2020pre} as the dominant form of nonlinearity. Hence, our proposed $\lambda_R$ range provides additional evidence supporting the dominant role of latitude quenching .

The variation of large-scale magnetic power around cycle minimum, especially the contrasting behavior of odd and even modes, such as the overestimation of the $l$=3 and $l$=5 modes and the underestimation of the $l$=2 mode with Profiles 1 and 2, offers a new perspective for understanding surface flux transport and constraining transport parameters. These results suggest that the poleward flow above about $\pm60^\circ$ latitudes is likely very weak or nearly absent. However, the non-axisymmetric mode gradually becomes dominant and makes the magnetic configuration more complex with the increase of solar activity. This requires more investigation to understand the impact of different transport processes on large-scale magnetic power during the cycle maximum. 

We note that the meridional flow profiles adopted in this study are primarily based on magnetic feature tracking, which was widely used in recent SFT studies. Compared with Doppler-shift measurements \citep{Ulrich2010} and helioseismology \citep{Zhao2014, Liang2018}, these measurements tend to yield slower flows at low latitudes and faster flows at higher latitudes. Comparing the magnetic power at $l \leq 5$ obtained from SFT models with different meridional flow profiles provides an effective method for constraining the realistic flow profile, which we plan to investigate in future studies.

The spectral power at $20 \lesssim l \lesssim 60$ is primarily determined by the assimilated AR sources, which are highly dependent on the AR identification and subject to inherent uncertainties. For example, small ARs that emerge on the far side of the Sun often appear highly diffused in synoptic maps and are therefore excluded from AR databases \citep{Wang2023, Wang2024}. Additionally, as noted by \cite{Wang2020}, unipolar regions resulting from the decay of earlier ARs introduce intrinsic challenges in identifying and quantifying subsequent ARs within activity complexes. These factors contribute to flux misidentification in synoptic maps and likely account for deviations between the observed and simulated power spectra at $20 \lesssim l \lesssim 60$.  
 
Although AR misidentification may affect the input magnetic flux, our results suggest that reasonably assimilated, realistic AR configurations can reproduce the spectral power at intermediate scales ($20 \lesssim l \lesssim 60$). Furthermore, when realistic transport parameters are applied, the power at larger scales ($l \lesssim 20$) can also be captured. However, most previous SFT models employed the BMR approximation for AR sources. Limitations of the approximation, particularly its impact on the axial dipole field, were discussed in the Introduction. The influence of the BMR approximation on the magnetic power spectra will be investigated in the subsequent paper in this series.


\begin{acknowledgments}
We thank the anonymous referee for the valuable comments and suggestions on improving the overall quality of this paper. The research is supported by the National Natural Science Foundation of China (grant Nos. 12425305, 12173005, and 12350004). SOHO is a project of international cooperation between ESA and NASA.
\end{acknowledgments}






\bibliography{power_spectra}{}
\bibliographystyle{aasjournal}



\end{document}